\newif\ifieeetran
\ieeetrantrue

\newif\ifbiography
\biographyfalse

\ifieeetran
	\documentclass[journal,a4paper]{IEEEtran}
\else
	\documentclass[12pt,onecolumn]{IEEEtran}
	\usepackage{setspace}
	\doublespace
	\usepackage[top=3.2cm, bottom=3.2cm, left=2.25cm, right=2.25cm]{geometry}
\fi

\newcommand{\ignore}[1]{}

\ifCLASSINFOpdf
\else
\fi
\usepackage{array}

\usepackage[tight,footnotesize]{subfigure}
\usepackage{url}

\usepackage{algorithm}
\usepackage{algpseudocode}
\usepackage{listings}
\usepackage{amsfonts}
\usepackage{graphicx, subfigure}
\usepackage{amssymb}
\usepackage{amsmath}
\usepackage{longtable}
\usepackage{tabularx}
\usepackage{multirow}

\usepackage{todonotes}
\usepackage[utf8]{inputenc}

\usepackage{fancyhdr}
\usepackage{fancyhdr}
\fancypagestyle{empty}{
\fancyhf{}

\lfoot{\parbox[]{0.95\textwidth}{\MakeLowercase{
\MakeUppercase{T}his article has been accepted for publication in a future issue of \MakeUppercase{IEEE}
\MakeUppercase{T}ransactions on \MakeUppercase{V}ehicular \MakeUppercase{T}echnology. \MakeUppercase{C}itation information: \MakeUppercase{DOI 10.1109/TVT.2016.2608942, IEEE}
\MakeUppercase{T}ransactions on \MakeUppercase{V}ehicular \MakeUppercase{T}echnology. \MakeUppercase{A} preprint version of this manuscript is available at: http://ieeexplore.ieee.org/document/7565650/}}}}
\usepackage{textcomp}
\begin{document}
%
\title{Modeling and Dimensioning of a Virtualized MME for 5G Mobile Networks}

%
%
%
\author{Jonathan~Prados-Garzon, Juan~J.~Ramos-Munoz, Pablo~Ameigeiras, Pilar~Andres-Maldonado, Juan~M.~Lopez-Soler
\thanks{
}
\thanks{
Copyright (c) 2015 IEEE. Personal use of this material is permitted. However, permission to use this material for any other purposes must be obtained from the IEEE by sending a request to pubs-permissions@ieee.org. 
}
\thanks{The authors are with the Department of Signal Theory, Telematics and Communications of the University of Granada, Granada, 18071 Spain (e-mail: jpg@ugr.es, jjramos@ugr.es, pameigeiras@ugr.es, pam91@correo.ugr.es, juanma@ugr.es).}
\thanks{Manuscript received [date]; revised [date].}}

\maketitle

\begin{abstract}
Network Function Virtualization is considered one of the key technologies for developing the future mobile networks. 
  In this paper, we propose a theoretical framework to evaluate the performance of an LTE \emph{virtualized Mobility Management Entity} (vMME) hosted in a data center. This theoretical framework consists of i) a queuing network to model the vMME in a data center, and ii) analytic expressions to estimate the overall mean system delay and the signaling workload to be processed by the vMME. We validate our mathematical model by simulation.
One direct use of the proposed model is vMME dimensioning, i.e., to compute the number of vMME processing instances to provide a target system delay given the number of users in the system. Additionally, the paper includes a scalability analysis of the system. In our study we consider the billing model and a data center setup of Amazon Elastic Compute Cloud service, and estimate experimentally the processing time of MME processing instances for different LTE control procedures. For the considered setup, our results show that a vMME is scalable for signaling workloads up to 37000 LTE control procedures per second for a target mean system delay of 1 ms. The database performance assumed imposes this limit in the system scalability.
\end{abstract}



\section{Introduction}
\label{sec:introduction}

 Nowadays, the telecom industry is considering Network Virtualization as one of the key technologies in the future 5G cellular networks. Network Functions Virtualization (NFV) offers the possibility of running the network functions on industry standard high volume servers (so-called commodity hardware) instead of using expensive, special purpose, and vendor-dependent hardware \cite{nfv-etsi}\cite{nfv-survey}. The decomposition of a service in a set of Virtual Network Functions (VNF) which can be executed in standard servers, allows for instantiating these VNFs in different network locations as needed. 
 Concretely, NFV promises to enable organizations to: i) reduce capital and operational expenditures, ii) accelerate time-to-market of new services, iii) deliver agility and flexibility, and iv) scale up services on demand \cite{nfv-etsi}. 

By way of illustration, nowadays the cellular networks are over-dimensioned in order to face the expected increase in the traffic load for the next years and considering the peak hours. The network entities are statically deployed and configured. Hence, there is a lack of network elasticity to deal with highly dynamic traffic patterns that might result in a waste of resources. Since NFV paradigm allows to create and scale network components on-demand, it can put an end to this problem. With the adoption of NFV, the mobile operators could adapt and optimize their resources in accordance with the given traffic conditions. 





This work aims at performing a dimensioning and scalability analysis of an LTE virtualized Mobility Management Entity (vMME) in a data center. On the one hand, the purpose of the vMME dimensioning is to determine the minimal number of processing instances required at the data center so that a given target mean system response time can be guaranteed.
Please note that dynamic resource provisioning is not addressed in this work, though the dimensioning may be part of such algorithms \cite{Pompili15}.
On the other hand, the scalability analysis of the vMME is intended to assess the productivity of the system, which depends on its running costs and performance in terms of throughput and delay.

In order to achieve these goals, in this paper we develop  a mathematical framework to assess the mean system response time of a vMME given the control messages arrival rate and the system service rates.
It will allow for estimating the time for a control message to be serviced. Our approach considers an 1:N mapping VNF implementation for the vMME \cite{Taleb15}. Using this architectural option, the vMME processing instances are stateless facilitating the resources scaling, high availability, and load balancing. 
Since this implementation design follows the multi-tiered web services deployment scheme for cloud-based applications \cite{Taleb15}, we use a queuing network similar to \cite{Vilaplana2014} and \cite{Lama13} for modeling the vMME in a data center. 
  In addition, we also provide analytic expressions to estimate
  the signaling workload to be processed by the vMME. We validate all this mathematical framework by simulation.


  The main contributions of this paper are the following:
\begin{itemize}

  \item The proposal of a theoretical model to compute the system response time of a vMME running in a data center.

\item A theoretical characterization 
 of the signaling workload generated by both the user's activity and the  Machine-Type Communications (MTC) devices.
We provide mathematical expressions to compute the rates of the LTE control procedures that generate most signaling load.

\item Using our theoretical framework we perform dimensioning of a vMME. 
We verify by simulation that the proposed framework is useful for that use case. This might be a first step in designing dynamic resource provisioning algorithms \cite{Pompili15}. 

 \item We also provide a scalability study of a vMME to investigate the evolution of the system productivity when it scales up its resources. In this work we consider the productivity as a metric that relates the performance, in terms of throughput and delay, and the running costs of the system. 

\end{itemize}
  
 In our study we consider both Human-Type Communications (HTC) and Machine-Type Communications (MTC). 
We use the data center setup and billing model of Amazon Elastic Compute Cloud (EC2). Based on this, we estimate experimentally the servicing rates of vMME processing instances for different LTE control procedures. We evaluate the vMME model by means of simulations considering the dense urban information society scenario of the METIS project \cite{metis} for 5G networks. We carried out the vMME scalability analysis and our results show that a vMME is scalable for signaling workloads up to 37000 LTE control procedures per second for a target average system delay of 1 ms. This limit in the system scalability is imposed by the database performance considered. 

 The paper is organized as follows. The next section provides some background and summarizes
relevant literature. Section \ref{sec:system-model} presents the system model. Section \ref{sec:control-procedures} reviews the LTE standard control procedures that are considered in this work. Section \ref{sec:traffic-models} describes the adopted traffic models.
In Section \ref{sec:modeling}, we propose the queuing model for the vMME, derive analytic expressions to compute the signaling workload and the mean system delay. This section also includes a simple vMME dimensioning algorithm.
Section \ref{sec:Scalability-metrics} provides a theoretical analysis to assess the virtualized MME scalability.
The system is simulated and evaluated in Section \ref{sec:numerical-results} where the proposed theoretical models are also validated. Finally, Section \ref{sec:conclusions} draws the main conclusions of the paper.

%

 
 
 
 

%

\section{Background and Related Works}
\label{sec:relatedworks}

The Mobility Management Entity (MME) is the key control entity for the LTE EPC. It interacts with the evolved NodeB (eNodeB), Serving Gateway (S-GW), and Home Subscriber Server (HSS) within the EPC to realize functions such as Non-Access Stratum (NAS) signaling, user authentication and authorization, mobility management (e.g. paging, user tracking), and bearer management \cite{3gpp-rel13}, among others.

Traditionally, MME was dimensioned to cope with the signaling workload expected for next years and considering the busy hours. Once the MME capacity was close to its limit (e.g., CPU load of 70\%), its hardware was upgraded to meet future needs while maintaining the same software and architectural design. One of the main drawbacks of the traditional approach is the lack of elasticity. To put an end to this issue, MME can leverage NFV paradigm to scale up and down depending on the current signaling workload. Furthermore, NFV makes viable the adoption of distributed architectures for MME \cite{Xueli12} by reducing provisioning costs.

Since NFV promises to bring substantial benefits to forthcoming mobile networks, there exists an intensive research work focused on this topic. There are works that have tackled the architectural and implementation issues of NFV in LTE Evolved Packet Core (EPC).
In \cite{Hawilo14} the authors discuss the challenges and requirements imposed by the adoption NFV paradigm in mobile networks. Furthermore, they propose an NFV framework for EPC and suggest a regrouping of its VNFs in order to reduce the control signaling. The authors in \cite{Taleb15} demonstrates that the implementation of EPC over a cloud infrastructure and providing it "as a Service" is feasible. They also present different architectural options and carry out a thorough analysis comparing these options.
The reference \cite{Takano14} describes a scheme for virtualization-based scaling of stateful network entities without interrupting user session continuity. Following this scheme and in order to prove its benefits, the authors design and implement an LTE virtualized Mobility Management Entity (vMME).  
The authors in \cite{Baba15} design and implement an architecture of a virtualized EPC tailored to the needs of the machine-to-machine services. They probe that their architecture proposal reduces CPU time consumption by up to 27\% by reducing control message volume.

Others works have addressed the study of the feasibility of the virtualization of the EPC. 
  For instance, the authors of \cite{Hirschman15} implement an entire EPC in general purpose processors. They argue the improved use of computational resources provided by NFV paradigm and show that servicing the synthetic workload generated by 50000 users is viable. 
   In \cite{rajan15}, the authors point out potential bottlenecks of a virtualized EPC (vEPC). To that end, they combine experimentation and analysis to demonstrate that the control plane signaling may severely interfere with the user plane packet processing. 

As far as the NFV-based applications are concerned, Project Clearwater \cite{clearwater} is an open source implementation of the IP Multimedia Subsystem (IMS) standard. Clearwater proposes a cloud-oriented design tailored for deployment in NFV ecosystem, which claims to be massively scalable and exceptionally cost-efficient. It makes use of stateless load balancing, which allows all components scaling out horizontally.

Regarding the modeling of virtualized networks, the author in \cite{Qiang11} proposes a model for service capabilities of composite network-Cloud service provisioning systems. Using deterministic network calculus, the aforementioned paper models these systems considering Latency-rate profile for the service components and a leaky bucket shaper to conform the user data traffic. 

\section{System Model} 
\label{sec:system-model}

In this work, we assume a general access cellular network architecture based on NVF, which also supports mobility and MTCs. Although this architecture reuses the entities defined in LTE/EPC, we simplify them to allow its extension to other cellular architectures. 



 
 
\begin{figure}[tb]
\begin{center}
\includegraphics[width=0.8
\columnwidth]{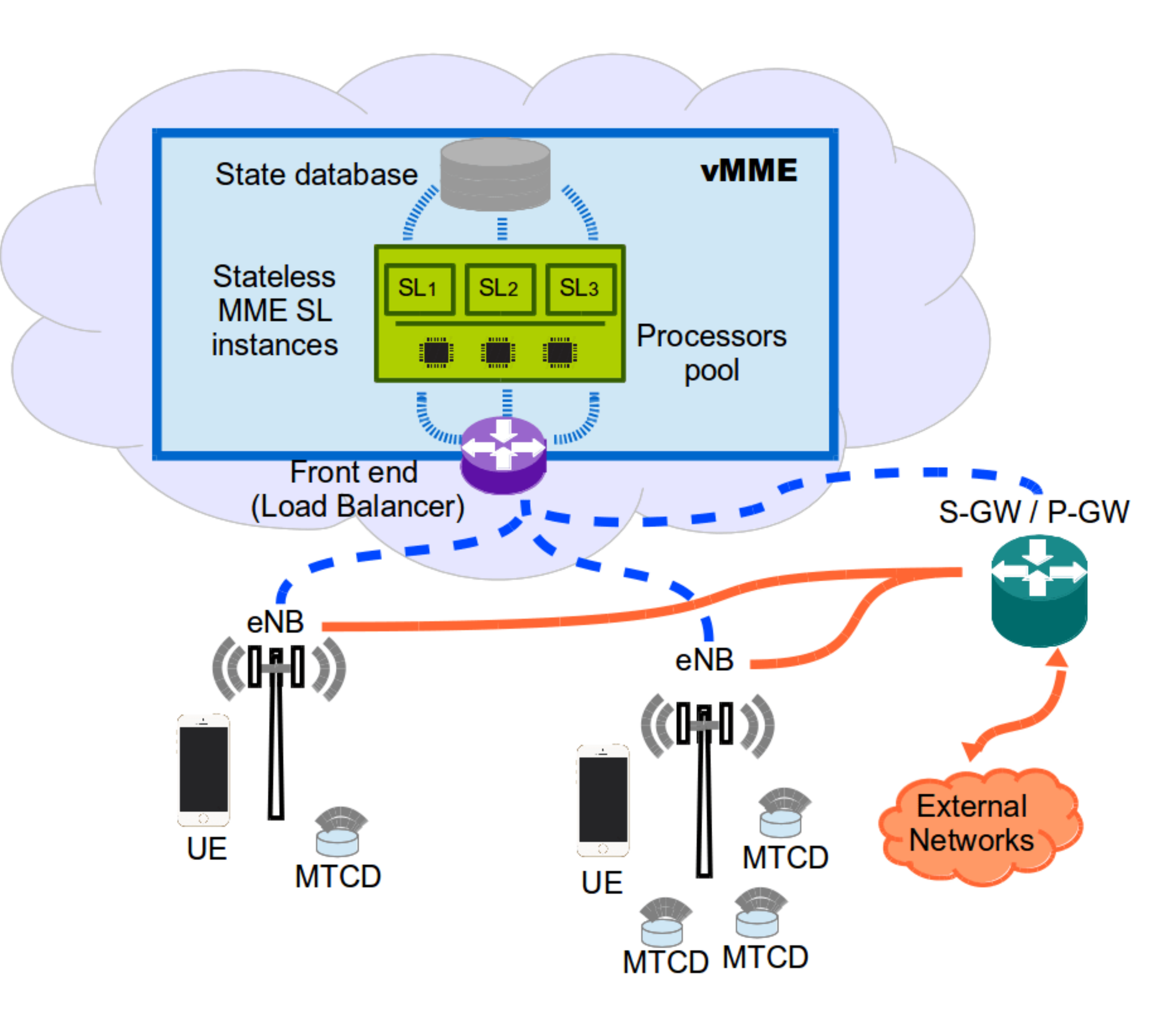}
\end{center}
\caption{Overall system model.}
\label{fig:5g-arch}
\end{figure}

 The overall system considered in this work is depicted in Fig. \ref{fig:5g-arch}. The main entities are explained next.
 
\subsection{The User Equipment (UE)}
 Let $N_U$ be the number of UEs in our system. UEs are the terminals which allow each user to connect to the network via the eNodeB base stations. We assume that UEs move following a fluid-flow mobility model. 
 The UEs run the users' applications which generate or consume network traffic, as described in Sec. \ref{sec:traffic-models}. 
    The UE is able to initiate requests to the network by using control messages. The activity of the UE and the generation of network traffic also trigger the network control procedures. 
    
       
\subsection{MTC devices (MTCDs)}  
 Let $N_D$ denote the number of MTCDs in the system. We assume the following assumptions: MTCDs are placed in fixed locations, they send small data packets to centralized servers infrequently, and additionally they use the same procedures as the UEs do to send their data. 
 
 \subsection{eNodeB stations (eNB)} They receive the UEs signaling and forwarding messages to the vMME. Each eNB contains a \emph{user inactivity timer} with an expiration time of $T_{I}$. Using this timer, the eNB detects the users' inactivity (i.e., the user does not perform any data communication over a period of length $T_I$) and can release network resources.

   \subsection{The virtualized Mobility Management Entity (vMME)} 
   The vMME is the main control entity of the network. It is in charge of maintaining the mobility state of the UE,  bearer management, and user authentication and authorization, among other functions. To support this functionality, LTE standard defines several signaling procedures (i.e., NAS procedures), which imply an exchange of signaling messages between the vMME and other LTE entities (e.g., eNB, S-GW and HSS). When the vMME receives one signaling message, it processes it, and later the vMME sends a new message to the another entity (such as eNB or S-GW). If the procedure requires several steps, the entity sends another response message to the vMME. Let $T_{IM}$ be the time between the vMME sends a control message to other LTE entity and the response message arrives at the vMME from that entity, where applicable. This time models the network delays and processing delay of the entity interacting with the MME.

   As far as the vMME implementation is concerned, we consider the \emph{1:N mapping} architectural option \cite{Taleb15}. Thus, the vMME is split into 3 logical components: front-end (FE), MME service logic (SL), and state database (SDB). The FE acts as the communication interface with other entities of the network and balances the load among several MME SLs, which implement the processing of the different control messages. In this way, the vMME is seen as a single component by the rest of the network. The SDB stores the user session state making the MME SLs stateless. 

   With regard to the operation between SDB and SLs, we assume that when an MME SL instance finishes processing a control plane message, it saves the transaction state and or the updated user context into the SDB. When a subsequent request arrives at an MME SL instance, it first gathers the user context (e.g., for deciphering the message) and transaction state from the database to continue from. The user context consists of a set of information elements associated with the user that can be categorized into user ID, user Location, Security, and EPS Session/Bearer information \cite{handoverX2netmanias}. As an example, let us consider the last message of a Handover (HR) procedure to be processed by the vMME. When an MME SL finishes processing this message, it will have to update some information of the user context (e.g., eNB UE S1AP ID, E-UTRAN Cell Global Identifier, and S1 Tunnel Endpoint Identifier for downlink) in the SDB.


This differs from vMME implementation based on Elastic Core Architecture \cite{Takano14}, and it allows fully stateless MME SLs. Different messages of the same procedure for the same user can be processed by different MME SL instances. Therefore, the number of MME SL instances, denoted as $m$, can grow without affecting on in-session users.

   When the processing capacity assigned to the vMME cannot withstand with the current control load, a new MME SL instance must be instantiated and a new processor is added to the processing resources pool. We presume \emph{dedicated hosting}, i.e., each MME SL instance runs on a server or a subset of servers and a server is allocated to at most one MME SL instance at any given time \cite{Urgaonkar05}. For simplicity, we will assume that every processor in the data center facility provides the same computational power. Moreover, we  consider that there are a single SDB and FE instances. 
\section{Control Plane Procedures}
\label{sec:control-procedures}

There exist several signaling procedures in LTE that allow the control plane to manage the UE mobility and the data flow between the UE and \emph{Packet Data Network Gateway} (P-GW). From all of them, we only concentrate on the ones that generate most signaling load \cite{Hirschman15}. 


In the following subsections, we describe the processing carried out by the MME during the control plane procedures that are considered in this work \cite{lte-signaling}.

 
 \subsubsection{Service Request (SR)}
 When a UE does not have  available resources and new traffic is generated, either from this UE or from the network to this UE, the UE performs a Service Request (SR) procedure. We focus on the UE-triggered SR. During this procedure the MME receives three different messages: an Initial UE Message ($SR_1$), an Initial Context Setup Response ($SR_2$), and a Modify Bearer Response ($SR_3$).
 
 
 To process the Initial UE Message ($SR_1$) the MME has to carry out UE integrity check and message decrypting. Additionally, it generates identifiers for the bearers to be established. Moreover, it stores and retrieves parameters and variables related to the UE context. Some of them are included in the subsequent Initial Context Setup Request message. During the processing of the Initial Context Setup Response message ($SR_2$), the MME also retrieves information of the UE context and includes this information in the subsequent Modify Bearer Request message. The processing of the Modify Bearer Response ($SR_3$) is minimum as this message is only a confirmation.
 
  
  

 \subsubsection{Service Release (SRR)}
 
 The Service Release (SRR) procedure is triggered by user inactivity. Its purpose is to release data radio bearers and downlink S1 bearer in the data plane, and radio and S1 signaling connections in the control plane for a UE. 
During the SRR, the MME processes three messages: a UE Context Release Request ($SRR_1$), a Release Access Bearers Response ($SRR_2$), and a UE Context Release Complete ($SRR_3$).
 

To process both the UE Context Release Request message ($SRR_1$) and the Release Access Bearers Request ($SRR_2$), the MME needs to retrieve information of the UE context and include this information in the subsequent messages. The processing of the UE Context Release Complete message ($SRR_3$) mainly implies the deletion of the bearer's context information by the MME.

  

 

 \subsubsection{X2-Based Handover (HR)}

The MME participates in the X2-based Handover (HR) during the handover completion phase. Its purpose is to switch the bearers' end point from the source to the target eNB. The MME receives two messages during this phase: a Path Switch Request message ($HR_1$) and a  Modify Bearer Response ($HR_2$).


To process both the Path Switch Request message ($HR_1$) and the Modify Bearer Response ($HR_2$), the MME also needs to retrieve information of the UE context and include this information in the subsequent messages. To process the Path Switch Request message, the MME also needs to store new information such as the IDs of the new serving cell and new tracking area.

\section{Traffic models}
\label{sec:traffic-models}

This section describes the traffic models considered in this work along with their statistical characterization. 

\subsection{Human type communication (HTC) traffic model}
\label{subsec:human-traffic-models}

\renewcommand{\tabularxcolumn}[1]{>{\arraybackslash}m{#1}}
\begin{table*}[tb]
\centering
\caption{Traffic models characterization}
\label{tab:traffic-models}
{\small
\begin{tabularx}{\textwidth}{c|c|l|X}
\hline
\hline

Com. Type & Traffic Type	& Parameters	& Statistical  Characterization \\                                              

\hline
\hline
\multirow{12}{*}[-13pt]{\parbox[c]{1.5cm}{\centering HTC \\ ($\overline{IAST} = 1200$ s \\ \cite{Ilias14})}} 

& \multirow{6}{*}[-6pt]{\parbox[c]{1.7cm}{\centering Web \\browsing \\(HTTP) $P_{app}=0.74$}}
 
 & Main Object Size	& Truncated Lognormal Distribution: $\mu$=15.098 $\sigma$=4.390E-5 min=100Bytes max=6MBytes 	
\\ \cline{3-4}                                                                            
& & Embedded Object Size	& Truncated Lognormal Distribution: $\mu$=6.17 $\sigma$=2.36 min=50Bytes max=2MBytes
\\
\cline{3-4}

& & Number of Embedded Objects per Page & Truncated Pareto Distribution: mean=22 shape=1.1
\\
\cline{3-4}

& & Parsing Time                        & Exponential Distribution: mean=0.13seconds
\\
\cline{3-4}

& & Reading Time                        & Exponential Distribution: mean=30seconds
\\ 
\cline{3-4}

& & Number of pageviews per session     & Geometric Distribution: p=0.893 mean=9.312
\\ 
\cline{3-4}

\cline{2-4}

 & \multirow{4}{*}{\parbox[c]{1.7cm}{\centering HTTP \\progressive \\video $P_{app}=0.03$}} 
 
 & Video Encoding Rate                 & Uniform distribution with ranges:  $(2.5, 3.0)$Mbps / (4.0,4.5)Mbps / $(12.5, 16.0)$Mbps / $(20.0,25.0)$Mbps, for equiprobable itags: 137 / 264 / 266 / 315 respectively. 
\\ 
\cline{3-4}

& & Video Duration & Distribution extracted from \cite{Ameigeiras12}                             \\ 
\cline{3-4}
                                        
& & Reading Time & Exponential Distribution: mean=30seconds                                  \\ 
\cline{3-4}
                                      
& & Number of video views per session & Geometric Distribution: p=0.6 mean=2.5                                        \\ 
\cline{3-4}

\cline{2-4}

 & \multirow{2}{*}{\parbox[c]{1.9cm}{\centering Video calling $P_{app}=0.23$}}         
 
 & Call Holding Time & Pareto Distribution: k=-0.39 s=69.33 m=0                                             \\ 
\cline{3-4}

& & Number of calls per session & Constant = 1 
\\  \cline{3-4}

\hline

\multirow{3}{*}[-10pt]{\parbox[c]{1.5cm}{\centering MTC}} 
& \multirow{3}{*}[-3pt]{\parbox[c]{1.6cm}{\centering Infrequent \\small data \\transmissions\\(Packet Size = 100 B)}}
 & Discretization time interval	& $\Delta_{T}=1$ sec
\\
\cline{3-4}

 & & Markov chain state transition matrix	& $P=\left( \begin{array}{cc}
1-p & q \\
p & 1-q \\
\end{array} \right)$ where $p=6.75\times 10^{-5}$ and $q=1.47\times 10^{-4}$
\\
\cline{3-4}

 & & Markov chain state rates	& $\lambda_{1}=0.0015$ packets/s; $\lambda_{2}=0.065$ packets/s
\\
\cline{3-4}

\hline \hline                                                 
\end{tabularx}
}
\end{table*}

\begin{figure}[tb]
\begin{center}
\includegraphics[width=1\columnwidth]{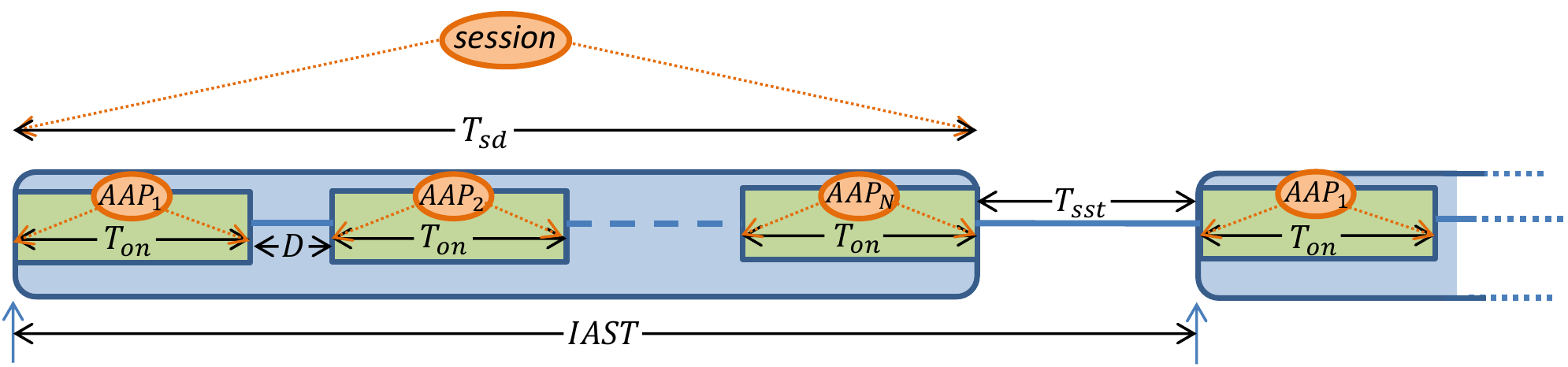}
\end{center}
\caption{HTC traffic model.}
\label{fig:HTC_model}
\end{figure}

Let us define a \emph{session} as the user activity 
elapsed between the instant the user launches a network application and the time instant he closes or stops it (Fig.  \ref{fig:HTC_model}). 


Likewise, \emph{Application Activity Period} ($AAP$) is defined as the time interval in which the application sends or receives all necessary data to perform a single task, such as download a web page, stream a video, or make a call. UE applications generate or consume traffic during the \emph{application activity periods} of a \emph{session}.

A session consists of $N$ $AAPs$ of length $T_{on}$ separated by $N-1$ \emph{reading times}. A \emph{reading time}
(of length $D$) is the time period elapsed between two successive activity periods within the same \emph{session}. During the \emph{reading time}, the user does actions such as reading the downloaded web page or deciding the next video to watch. 

Let us define \emph{Inter-Arrival Session Time} ($IAST$) as the time interval between the start of two consecutive sessions. And let $T_{sst}$ denote the \emph{session standby time}, i.e., the time elapsed from the end of a session to the beginning of the next one.

We assume $T_{sst}$ follows an exponential distribution with mean  $\overline{T}_{sst}=(\overline{IAST} - \overline{T}_{sd})$ seconds, where $T_{sd}$ is the session duration and
 $\overline{A}$ denotes $E[A]$, for any $A$. 

Assuming that $N$, $D$ and $T_{on}$ are statistically independents, it holds that:
\begin{equation} \overline{T}_{sd}=\overline{N} \cdot \overline{T}_{on} + (\overline{N} - 1) \cdot \overline{D}. 
\end{equation}

 Whenever a session begins, the user chooses a certain application with a given probability $P_{app}$ (see Table \ref{tab:traffic-models}). Three types of applications are considered in this work: i) web browsing, ii) HTTP progressive video and iii) video calling. 
The specific values of $P_{app}$ used for each application were computed from the percentages of total network traffic generated per type of traffic given in TC2 scenario of the METIS project \cite{metis}.
Similarly, to estimate the data rates of the future mobile traffic, we have followed the predictions assumed in the METIS project \cite{metis}.
Table \ref{tab:traffic-models} summarizes the statistical characterization of the considered application models, which are briefly described below.

\subsubsection{Web Browsing}

 The characterization of this traffic is described in \cite{ngmn08}. The amount of data downloaded for an \emph{application activity period} (i.e., web page size) of a web browsing session is determined by the main object size (i.e. the HTML file), the number of embedded objects and their sizes. During a session, the number of downloaded web pages per session is set to follow a geometric distribution
 \cite{Gou11}.

 The download time is determined by the web page size, the link data rate, and the \emph{parsing time}. The \emph{parsing time} is defined as the time interval the web browser takes to parse the embedded objects. 

 We estimate the future web pages sizes by extrapolating the data series of \cite{httpArchive}, and scaling main objects size, accordingly. 
 
\subsubsection{HTTP progressive video}

 For this type of traffic, we adopt the YouTube model of \cite{Ameigeiras12}, in which a video is transferred at a constant and limited rate during a \emph{throttling phase} after an initial period of high downloading rate, called \emph{initial burst}. The number of downloaded video clips per session is set to follow a geometric distribution
 \cite{Phillipa08}.  We assume that the \emph{reading times} for  this model and for web browsing are identically distributed.
 
 The size of each video is calculated from its duration and encoding rate. The video encoding rate depends on the video format selected. Each video format, identified by an \emph{itag} number, determines a container file format, an encoding algorithm, and a video resolution. To meet the METIS predicted data rates, we have considered the YouTube video formats with the highest encoding rates and resolutions.  
 
 The video download time (i.e., activity period) is determined by the bottleneck link data rate during the initial burst and limited by the media server during the throttling phase \cite{Ameigeiras12}.
 

\subsubsection{Video calling}
 In this application, a session starts when the user opens a video calling client app and makes a single call to someone else. This application generates constant bit rate traffic at 1.5 Mbps which is the recommended download/upload speed of Skype for HD video calling.
 
 The call duration or \emph{call holding time} determines the application activity period duration. The statistical characterization for the call duration has been extracted from \cite{Trang04}.


\subsection{Machine type communication (MTC) traffic model}
\label{subsec:machine-traffic-models}
In this work we implement the MTC traffic model based on Markov-modulated Poisson processes (MMPPs) from \cite{m2mbook}, but without taking into consideration the coordinated behavior for MTC devices.

In this model, each MTC device with index
$j=\{1,2,...,N_{D}\}$ is modeled by an MMPP. 
Let us define $n$ as the time index resulting of time discretization $n=\frac{t}{\Delta_{T}}$ for any constant time interval $\Delta_{T}$. An MMPP is a Poisson process modulated by the rate $\lambda_{j}^{MTC}[n]$, which is given by the state of a Markov chain $s_{j}[n]$. Then, $\lambda_{j}^{MTC}[n]=\lambda_{i}$ when $s_{j}[n]=i$, where $i=\{1,2,...,I\}$ denotes the index of Markov state and $\lambda_{i}$ denotes a constant rate associated to the state $i$. 



Assuming a constant packet size of $100$ bytes for MTC devices, 
we use the parameters listed in Table \ref{tab:traffic-models} for this model. This setup is extracted from \cite{m2mbook}, which corresponds to 
a fleet management service case. 



\section{vMME Queuing Model}

\label{sec:modeling}



\subsection{Model description}
\label{sec:queuing-model}

 To model a vMME with a 1:N mapping architecture
 as described in Section \ref{sec:system-model}, we consider a queuing system based on \cite{Vilaplana2014} which  
models a typical cloud processing chain. We assume that all the MME SL instances have the same computation power. Table \ref{tab:queue-parameters} provides the notation and main definitions for describing the queuing system.
 
  
   
\begin{figure}[t]
\begin{center}
\includegraphics[width=1.0\columnwidth]{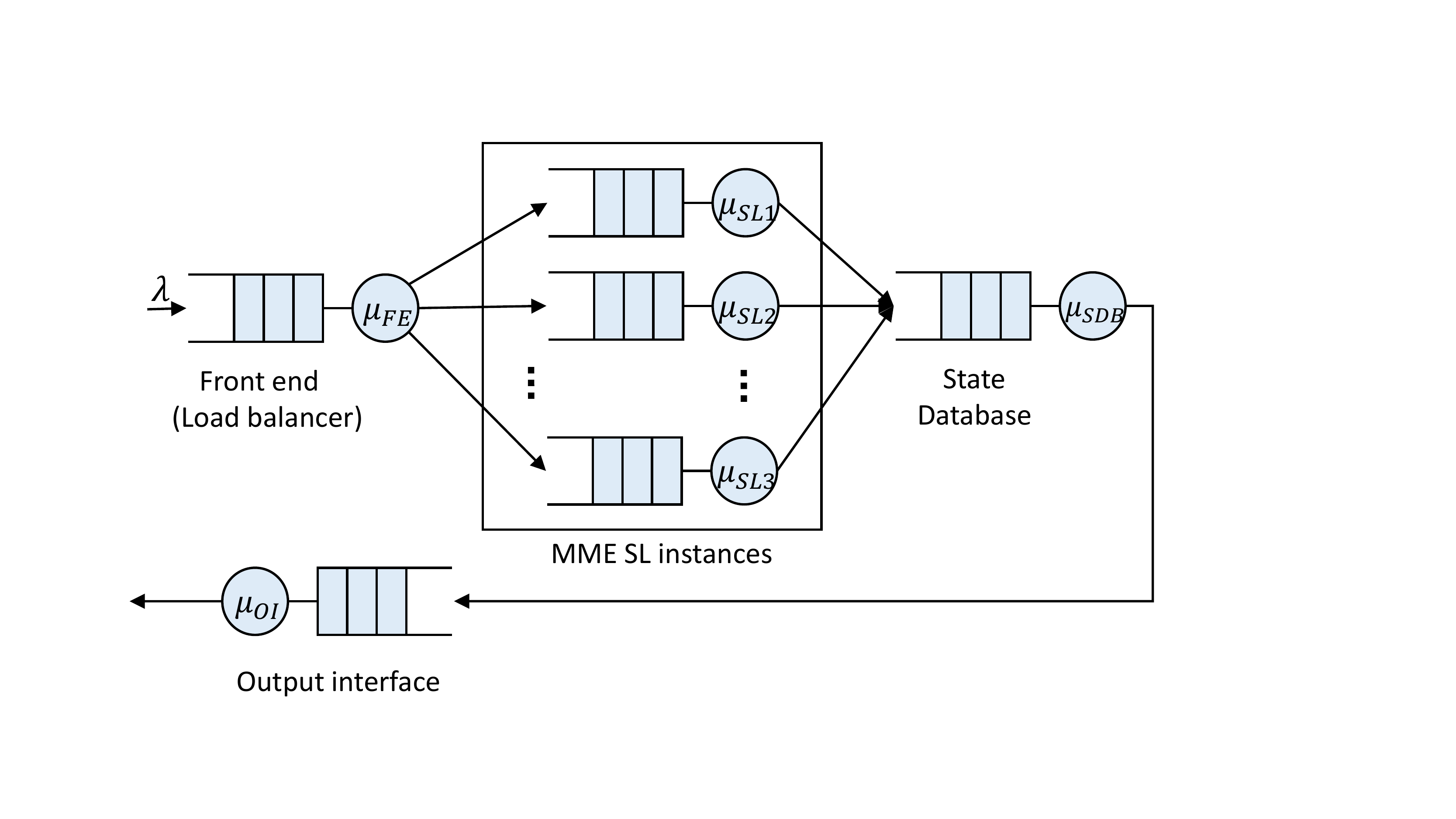}
\end{center}
\caption{vMME queue model.}
\label{fig:nfvqueue}
\end{figure}

 On the one hand, the state database, the FE, which balances the control requests among the MME SL instances, and the output network interface are modeled with single processor queues, with service rates respectively denoted by $\mu_{SDB}$, $\mu_{FE}$ and $\mu_{OI}$  (Fig. \ref{fig:nfvqueue}).
  On the other hand, the MME SL pool is modeled by a set of queues and processors that allow the parallel processing of the control messages.
  

 
  \begin{table}[tb]
\centering
\caption{Primary 
definitions.}
\label{tab:queue-parameters}
\begin{tabular}{ll}
\hline
\hline
Notation & Description \\
 \hline
 $D$     & Reading time length. \\
 $N$	 & Number of AAP per session. \\
 $N_{U}$ & Total number of HTC UEs. \\
 $N_{D}$ & Total number of MTC devices. \\
 m & Number of MME SL instances. \\
 $\lambda$ & Total arriving rate. \\
 $\lambda_{HR}$ & Arriving rate of \emph{Handover} requests. \\
 $\lambda_{SR}$ & Arriving rate of \emph{Service Request} requests. \\
 $\lambda_{SRR}$ & Arriving rate of \emph{Service Release Request} requests. \\
 $r_{cc}$ & Cell crossing rate. \\
 $T$ & Total system processing time. \\
 $T_{max}$ & Target mean system delay. \\
 $T_{DB}$ & Database processing time.\\
 $T_{FE}$&  Front-end processing time. \\
 $T_{SL}$ & MME SL processing time.\\
 $T_{on}$ & Duration of the AAP. \\
 $T_{OI}$ & Output interface processing time.\\
 $T_{sd}$ & Session duration. \\
 $T_{sst}$ & Duration of the session standby time. \\
 $\mu_{SDB}$ & Average database service rate.\\ 
 $\mu_{FE}$ & Average front-end service rate.\\ 
 $\mu_{OI}$ & Average output interface service rate.\\ 
 $\mu_{SL}$ & Average MME SL instance service rate.\\ 
 $O_{bw}$ & Egress link capacity.\\
 $O_{size}$ & Average response packet size.\\
\hline
\hline
\end{tabular}
\end{table}

\subsection{Arrival rate calculations for signaling requests}
\label{sec:arrival-rate}

 In this section, we derive mathematical expressions to predict the arrival rate of signaling procedure requests to the vMME. It will depend on the activity of the HTC UEs and MTCDs. Let $\lambda$ be the aggregate control messages arrival rate. 
 Then, from the description of the control procedures of Section \ref{sec:control-procedures}, $\lambda$ is calculated as
\begin{equation}
  \lambda=3\cdot \lambda_{SR}+3\cdot \lambda_{SRR}+2\cdot \lambda_{HR}
  \label{eq:lambda}
\end{equation}
  Where $\lambda_{SR}$, $\lambda_{SRR}$, and $\lambda_{HR}$ denote the mean arrival rates for SR, SRR and HR procedures, respectively. These rates can be expressed in terms of mean arrival rates per user $\lambda_{U}^{h}$ and per MTC device
  $\lambda_{S}^{h}$ for procedure $h$ $\in \{SR, SRR, HR\}$. Then, 
  \begin{equation}
  \lambda_{h} = N_{U} \cdot \lambda_{U}^{h} + N_{D} \cdot \lambda_{S}^{h}
  \end{equation}
  We suppose no mobility for MTCDs, thus $\lambda_{S}^{HR}=0$. We describe each arrival rate in the following paragraphs.   

\subsubsection{$\lambda_{U}^{SR}$ and $\lambda_{U}^{SRR}$ calculations}

 An SR procedure occurs whenever a UE is going to start an AAP without having network resources assigned. When an AAP finishes, a \emph{user inactivity timer}, whose value is denoted as $T_{I}$, starts. 
 Let $X$ denote the time elapsed between the end of an AAP and the beginning of the next one, regardless these activity periods belong to the same session or not. If $X\geq T_{I}$, the SRR procedure is triggered.

Since each SR has a corresponding SRR, it holds that
\begin{equation}
  \lambda_{U}^{SRR}=\lambda_{U}^{SR}
\end{equation}
Let $\overline{N}_{S}^{SR}$ be the mean number of SRs procedures per session, which is given by:
\begin{equation}
  \overline{N}_{S}^{SR}=\overline{N} \cdot P(X > T_{I})
\end{equation}
where $P(X > T_{I})$ is the probability that the inactivity timer expires.

 

 Finally, let us $\lambda_{S}=1/\overline{IAST}$ denote the sessions rate, i.e., the mean number of sessions per unit time. It holds that 
 \begin{equation}
 \lambda_{U}^{SR}=\lambda_{S} \cdot \overline{N}_{S}^{SR}
 \label{eq:rate_sr_ue_one}
 \end{equation}
 Since for the first activity period $X=T_{sst}$ and for the following $N-1$ ones $X=D$, then 
\begin{equation}{\lambda_{U}^{SR}=\lambda_{S}\cdot ((\overline{N}-1)\cdot P(D>T_{I}) + P(T_{sst}>T_{I}))}
\label{eq:lambdaSR}
\end{equation}



\subsubsection{$\lambda_{U}^{HR}$ calculation}

Assuming that each eNodeB serves only one cell, an HR procedure takes place when a user performs a cell change while being active. A user is considered active from the triggering of the SR procedure to the triggering of the associated SRR event. Let $P_{UA}$ denote the probability that a user is active at a given time, and let $\overline{r}_{cc}$ be the mean user cell crossing rate, i.e., the average number of cell crossings per unit time.
Thus, the mean arrival rates per user for HR is
\begin{equation}{\lambda_{U}^{HR}=\overline{r}_{cc}\cdot P_{UA}}
\label{eq:lambdaHR}
\end{equation}
 Assuming that each user moves according to the fluid-flow mobility model, i.e., at constant speed with random direction uniformly distributed between $[0, 2\pi)$, 
 it holds that
\begin{equation}
 \overline{r}_{cc}=\frac{\overline{v} \cdot B}{\pi \cdot S}
\label{eq:ccr}
\end{equation}
 where $\overline{v}$ is the mean user speed, and $B$ is the perimeter of the cell coverage area $S$. 
 
To compute $P_{UA}$, 
 let $T_{ua}$ denote the temporal extension
 of an AAP, defined as the time interval elapsed from the end of an AAP to the inactivity timer expiration or to the beginning of the next activity period, whichever comes first, that is, $T_{ua}=X$ if $X \leq T_{I}$ and $T_{ua}=T_{I}$ otherwise.
Then, $T_{ua}$ will follow the same distribution as $X$, but upper truncated to the value of $T_{I}$. Thereby, the expected value of $T_{ua}$ can be computed as
\begin{equation}{\overline{T}_{ua}(X)=T_{I}\cdot P(X>T_{I})+\int_{0}^{T_{I}} x\cdot f_{X}(x) \, dx}
\label{eq:ta}
\end{equation}
 Therefore, $P_{UA}$ is $\lambda_{S}$ times the amount of time that a user is active within a session:
\begin{equation}
 P_{UA}=\lambda_{S} \cdot (\overline{N}\cdot \overline{T}_{on}+(\overline{N}-1)\cdot \overline{T}_{ua}(D)+\overline{T}_{ua}(T_{sst}))
\label{eq:pa}
\end{equation}

\subsubsection{$\lambda_{S}^{SR}$ and $\lambda_{S}^{SRR}$ calculations} An SR procedure occurs whenever an MTC device is going to transmit a new packet without having network resources allocated. Let $P(t_{r}>T_{I})$ denote the probability that the time interval between two packets transmission for any MTC device $t_{r}$ be greater than inactivity timer value $T_{I}$. It holds that
\begin{equation}
\lambda_{S}^{SR}=P(t_{r}>T_{I})\cdot \sum\limits_{i=1}^{I} \lambda_{i} \cdot \pi_{i}
\label{eq:lambda_sr_mtc}
\end{equation}
where $\pi_{i}$ is the probability of the state $i$ and $I$ the number of states of the Markov chain. For our case $I=2$, $\pi_{1}=\frac{q}{p+q}$ and $\pi_{2}=\frac{p}{p+q}$.

Again, it verifies that
\begin{equation}
\lambda_{S}^{SRR}=\lambda_{S}^{SR}
\label{eq:lambda_srr_mtc}
\end{equation}

   
   



\subsection{vMME Response Time}
 
  To estimate the system response time, we suppose that there is a single Poisson arrival stream with arrival rate $\lambda$ which represents the control plane messages sent to the vMME (see Fig. \ref{fig:nfvqueue}), and that all the processing elements of the system are exponential servers with a service rate $\mu_{i}$ calculated from their mean service time $\overline{t}_i$ as $\mu_{i}=[\overline{t}_i]^{-1}$. Although these are strong assumptions, it allows us to assume a Jackson's open network, what eases obtaining analytical expressions. As we will show later in Section \ref{sec:numerical-results}, the proposed vMME analytic model provides a fairly good approximation to the values obtained by simulation.
  
 Let $\overline{T}_D$ denote the mean response time of the FE node and let $\overline{T}_{OI}$ denote the output 
 interface mean response time. Let $\overline{T}_{SL}$ denote the mean response time of the servicing nodes. And let $\overline{T}_{DB}$ denote the average processing time of the database.
 If the modeled system is assumed to be an open Jackson network, the mean response time $\overline{T}$ of the entire system of Fig. \ref{fig:nfvqueue} can be estimated by
  
\begin{equation}
 \overline{T}=\overline{T}_{FE}+\overline{T}_{SL}+\overline{T}_{DB}+\overline{T}_{OI}
   \label{eq:T}
\end{equation}

\subsubsection{$\overline{T}_{FE}$ calculation}: The front-end is modeled with an \emph{M/M/1} queue. Therefore, $\overline{T}_{FE}$ can be calculated as 
  \begin{equation}
   \overline{T}_{FE}=\frac{(\mu_{FE})^{-1}}{1-\lambda/\mu_{FE}}
   \label{eq:TIS}
  \end{equation}
  where the $\mu_{FE}$ is the mean service rate of the front-end node.
   
 \subsubsection{$\overline{T}_{SL}$ calculation}  The services nodes are modeled with an $M/M/m$ queue, and therefore their mean response time are computed as
  \begin{equation}
   \overline{T}_{SL}=\mu_{SL}^{-1}+\frac{C(m,\rho)}{m\cdot \mu_{SL} - \lambda}
   \label{eq:TNFV}
  \end{equation}
 where $\rho=\frac{\lambda}{\mu_{SL}}$, and $C(m,\rho)$ represents the Erlang's C formula calculated as
 \begin{equation}
   C(m,\rho)=\frac{\left( \frac{(m\cdot \rho)^m}{m!} \right) \cdot \left( \frac{1}{1 - \rho} \right)}
		  {\sum_{k=0}^{m-1} \frac{(m\cdot \rho)^k}{k!} + \left( \frac{(m\cdot \rho)^m}{m!}\right) \cdot \left( \frac{1}{1-\rho}\right)}
   \label{eq:cmrho}
  \end{equation}
   
 The average service rate of the NFV procedures, $\mu_{SL}$ is equal to $[\overline{t}_{SL}]^{-1}$. We estimate the average service time $\overline{t}_{SL}$, by weighting the processing time of each procedure according to its frequency, as explained in Section \ref{sec:control-procedures}. Consequently, 
\begin{align}
\overline{t}_{SL}=&\frac{\lambda_{SR}}{\lambda}\cdot (t_{SR_1}+t_{SR_2}+t_{SR_3})+ \nonumber \\
&\frac{\lambda_{HR}}{\lambda}\cdot(t_{HR_1}+t_{HR_2})+ \nonumber \\ 
&\frac{\lambda_{SRR}}{\lambda}\cdot (t_{SRR_1}+t_{SRR_2}+t_{SSR_3}) \label{eq:tnfv}
\end{align}
 where $t_{SR_i}$ is the processing time of the $i^{th}$ step of the Service Request procedure, $t_{SRR_i}$ is the processing time of the $i^{th}$ step of the Service Release Request procedure, and $t_{HR_i}$ is the processing time of the $i^{th}$ step of the Handover procedure. 

\subsubsection{$\overline{T}_{DB}$ calculation} The processing time of the database stage can be estimated by:
   \begin{equation}
    \overline{T}_{DB}=\frac{1/\mu_{SDB}}{1-\lambda/\mu_{SDB}}
  \label{eq:TDB}
  \end{equation}
    where $\mu_{SDB}$ is the service rate of the database server.
    
 
 \subsubsection{$\overline{T}_{OI}$ calculation}
   Finally, the output interface service rate $\mu_{OI}$ is calculated as $O_{bw}/O_{size}$, where $O_{bw}$ is the output link bandwidth and $O_{size}$ is the average packet size of responses. Therefore,
  \begin{equation} T_{OI}=\frac{(O_{bw}/O_{size})^{-1}}{1-(\lambda)/(O_{bw}/O_{size})}
   \label{eq:TOS}
  \end{equation}
\subsection{vMME dimensioning criterion}
\label{subsec:dimensioning-criterion-theo}

 The analytic proposed model is useful for dimensioning of a virtualized MME. For example, it can provide the minimum number of MME SL instances $m$ needed to achieve a target mean system delay $\overline{T}_{max}$ for a given number of UEs $N_{U}$ and MTCDs $N_{D}$. Assuming the services rates $\mu_{SDB}$, $\mu_{FE}$, and $\mu_{OI}$ are high enough to withstand the required MME signaling load, it holds that:
\begin{equation}
   m=\min\{M : \overline{T}(\lambda, M) \leq \overline{T}_{max}, M \in  \mathbb{N}\}
   \label{eq:m}
\end{equation}
Therefore, $m$ can be found with a simple algorithm that interactively increases the number of MME SL instances until it obeys that $\overline{T}(\lambda, m) \leq \overline{T}_{max}$.



%
\section{vMME scalability analysis}
\label{sec:Scalability-metrics}


 The virtualized MME proposed in this work is a distributed system. To complete the study of this system, we assess its scalability in this section. To that end, we adopt the scalability metric defined in \cite{Jogalekar00}. 
 This metric is based on productivity: the distributed system is scalable if the productivity is maintained as the system scale changes. 

 The scalability metric for one system at two different scale factors $k_2$ and $k_{1}$, noted as $\psi(k_{1},k_2)$, is defined as the ratio between the productivity of two systems at scale $k_2$ and $k_1$ \cite{Jogalekar00}:
 \begin{equation}
  \psi(k_{1},k_2)=F(k_2)/F(k_{1})
 \end{equation}
 where the productivity $F(k)$ represents the throughput delivered by the system over cost incurred per second for the scale factor $k$, calculated as: 
\begin{equation}
F(k)=\frac{\lambda(k) \cdot f(k)}{C(k)}
\end{equation}
 where $\lambda(k)$ denotes the average throughput attained at scale $k$, $C(k)$ denotes the running costs of the system at scale $k$, and $f(k)$ is a function of some appropriate system measures. In this work we use the $f(k)$ function defined in 
\cite{Jogalekar00}, which calculates the average response time $\overline{T}(k)$ compared to a target value $\widehat{T}$:
 \begin{equation}
  f(k)=1/(1+\overline{T}(k)/\widehat{T})
 \end{equation}



  





 Generally, for a given system, the scalability metric is defined in absolute terms and denoted by $\psi(k)$, since 
 the value of  $\psi(k_1,k_2)$ at $k_1$ is fixed at a known value. In this case,  $\psi(k)$ is interpreted as follows. If  $\psi(k)=1$, the system perfectly scales with $k$. If  $\psi(k)>1$, then the system scales positively with $k$. If  $\psi(k)<\gamma$, the system does not scale. In this work, we adopt $\gamma=0.8$ as in \cite{Jogalekar00}.

 
 

  


A strategy for scaling up the system is defined by the scaling factor $k$ and several scaling variables which depend on $k$.
In this work, we set as the scaling variable $m=k$. Therefore,  the reference scale factor $k_1$ corresponds to the system with one NFV instance. Additionally, for a given $k$ factor, the other system variables are configured to serve the maximum number of UE within a $T_{max}$ service delay budget. 


%
 To consider a realistic cost function for our vMME cloud-based system, we consider the Amazon EC2 Service billing model \cite{amazon-ec2}. To that end, let $C_{ci}(m)$ denote the per instance computing cost, let $C_{b}(m)$ denote the load-balancer service cost, and let $C_{db}(m)$ denote the database accessing cost. Then, the total cost $C(m)$ is 
 \begin{equation}
 C(m)=C_{b}(m)+ m \cdot C_{ci}(m)+C_{db}(m)
 \label{eq:ck}
\end{equation}
  where $m$ is the number of virtualized MME SL instances.
  Each element's cost includes a rental fee, a storage charge, and a per transaction or throughput price, as we describe next. The exact cost calculation depends on the cloud's billing model. To complete our study, Section \ref{sec:numerical-results} includes a numerical example that shows the practical applicability  of the conducted analysis.
  
  Cost $C_{ci}(m)$ includes a per unit time billing costs depending on the type of processor $C_{ci_{type}}(m)$, the cost of the outgoing traffic sent to Internet $C_{ci_{thro}}(m)$ per unit time, and the per computing instance storage cost $C_{ci_{stor}}(m)$:
\begin{equation}
C_{ci}(m) = C_{ci_{type}}(m)+C_{ci_{stor}}(m)+C_{ci_{thro}}(m)
\end{equation}
 The database accessing cost $C_{db}(m)$ includes a rental fee per unit time $C_{db_{type}}(m)$, the cost per data capacity $C_{db_{stor}}(m)$, and a fee per transactions per unit time $C_{db_{trans}}(m)$.
\begin{equation}
 C_{db}(m)=C_{db_{type}}(m) + C_{db_{stor}}(m)+ C_{db_{thro}}(m) 
\end{equation} 
The considered cloud service provides a load balancer service. Its cost $C_b(m)$ is charged by activation time $C_{b_{type}}(m)$ and served throughput $C_{b_{thro}}(m)$.
 \begin{equation}
 C_{b}(m)=C_{b_{type}}(m)+C_{b_{thro}}(m)
\end{equation} 

\section{Numerical results} 
\label{sec:numerical-results}


In this section, some numerical results are reported. It aims at validating the proposed mathematical framework to model a vMME and evaluating its scalability.


Our evaluation framework includes two software tools: a generator of procedure calls and a queuing system simulator.

The generator of procedure calls is implemented in the ns-3 simulator \cite{ns3}.
It implements the traffic models presented in Section \ref{sec:traffic-models} and the corresponding network signaling. The simulation scenario considered for each user is based on the dense urban information society scenario of the METIS project \cite{metis}. 
It is composed of $12$ eNodeBs distributed regularly in a $4 x 3$ grid over a rectangular area of size $387\,m\,x\,552\,m$. The coverage area for each eNB is rectangular with dimensions of $138\,m\,x\,129\,m$. The users move across the area following a fluid-flow mobility model.
The user speed is uniformly distributed between $0$ and $4.2\,m/s$.

The percentage of traffic generated for each type of application has been adjusted to meet the simulation guidelines of METIS project (see Table \ref{tab:traffic-models}) \cite{metis}. All users have an independent and constant uplink and downlink data rate of $300\,Mbps$ \cite{metis}. 
During the simulation, each control procedure 
generates control messages which are dumped to a trace file. 

The queuing system simulator implements the queuing model presented in Section \ref{sec:queuing-model} using the Matlab Simulink framework. 
The queuing model is fed with the traces produced by the previous tool. The load balancer has a service rate of $120000$ packets per second \cite{rightscal}. The database service rate has been obtained by assuming that the database deployed in the Amazon Cloud is the Amazon Aurora database \cite{amazon-ec2}, which is reported to serve $100000$ transactions per second \cite{aurora-benchmark-2015}. The output interface is a 10G Ethernet that serves up to $5000000$ packets per second (i.e., assuming an average packet size of 250 Bytes for the control messages generated by the vMME). 

In the Simulink model, we included an infinite server ($G/D/\infty$ queue), which models the one-way transmission delay and processing times of the network nodes from any eNB to the vMME. This delay was set to 7.5 ms. Another similar server was used to implement the parameter $T_{IM}$ (i.e., the time between the vMME sends a control message to other LTE entity and the response message arrives at the vMME from that entity). $T_{IM}$ was fixed to 15 ms.




 

\newcolumntype{P}[1]{>{\centering\arraybackslash}p{#1}}
\begin{table}[tb]
\centering
\caption{Parameters Configuration}
\label{ParametersConfiguration}
\begin{tabular}{|p{3.9cm}|P{4cm}|}
\hline
\multicolumn{2}{|c|}{RAN topology}  \\ \hline
eNBs layout & Regular Grid 387 m x 552 m    \\ \hline
eNB coverage area & 138 m x 129 m         \\ \hline
Number of eNBs  & 12 \\ \hline
\multicolumn{2}{|c|}{UE mobility}                        \\ \hline
Mobility model & Fluid-flow model                      \\ \hline
Speed  & Uniform distribution (0, 4.2) m/s \\ \hline
\multicolumn{2}{|c|}{EPC delays}  \\ \hline
One-way delay (eNB $\rightarrow$ vMME) & 7.5 ms \\ \hline
$T_{IM}$ (vMME $\rightleftharpoons$ [eNB $|$ S-GW]) & 15 ms \\ \hline
$\overline{T}_{max}$ & 1 ms \\ \hline
\multicolumn{2}{|c|}{Service rates}                        \\ \hline
FE service rate ($\mu_{FE}$)   & 120000 packets per second 
\\ \hline
SDB service rate ($\mu_{SDB}$)   & 100000 transactions per second 
\\ \hline
OI service rate ($\mu_{OI}$)   & 5000000  packets per second 
\\ \hline
\end{tabular}
\end{table}


\subsection{vMME processing time $(T_{SL})$ estimation
}

\label{sec:cpu-numerical-results}
In this section we calculate the MME SL processing times for each control message, what will be used as the MME SL average service time (and its respective service rate $\mu_{SL}$). Given a CPU processing capacity, we can estimate the delay of processing a message by assessing the average number of CPU instructions required for running a particular procedure.   
 
 To do this, we have considered the CPU characteristics of a real cloud service configuration from the \emph{Amazon Elastic Compute Cloud (EC2)} \cite{amazon-ec2}. Additionally, we have implemented in C the code of the functions which are invoked in the vMME for each procedure.
 
Although our implementation may differ from complete MME VNF implementations, we think that our version executes similar tasks as those ones. 
 
  After compiling the code, we measured the number of CPU instructions executed for every procedure by means of profiling tools.
The results drew
that the number of run instructions for the different control messages is very similar (see Table \ref{tab:procedure-response-time}).
 Table \ref{tab:procedure-response-time} also provides the delays
 calculated for the \emph{EC2 m3.xlarge} virtual instance of the Amazon EC2 service \cite{amazon-ec2}. The average computing capacity of this type of instance is
    $11.38\cdot 10^9$
  float operations per second \cite{iosup2011performance}.


   




\begin{table}[tb]
\centering
{\small
\caption{Number of Instructions and Processing Times in \emph{m3.xlarge} instance.}
\label{tab:procedure-response-time}
\begin{tabular}{|l|c|c|}
\hline
    Procedure &  \parbox[c]{2cm}{\centering Number of\\Instructions} & \parbox[c]{2,5cm}{\centering Processing\\Time, $T_{SL}$ ($\mu s$)}\\ 
\hline
\hline
$SR_1$ & 1.45e+06 &  127.4 \\ 
$SR_2$ & 1.07e+06 & 94.0 \\ 
$SRR_1$ & 1.07e+06 & 94.0 \\ 
$SRR_2$ & 1.07e+06 & 94.0 \\ 
$SRR_3$ & 1.06e+06 & 93.2 \\ 
$HR_1$ & 1.07e+06 & 94.0 \\ 
$HR_2$ & 1.07e+06 & 94.0 \\ 
\hline
\end{tabular}
}
\end{table}
\subsection{Signaling Procedures Rate}
\label{sec:arrival-rate-results}
\begin{figure}[tb]
\begin{center}
\includegraphics[width=1.0\columnwidth]{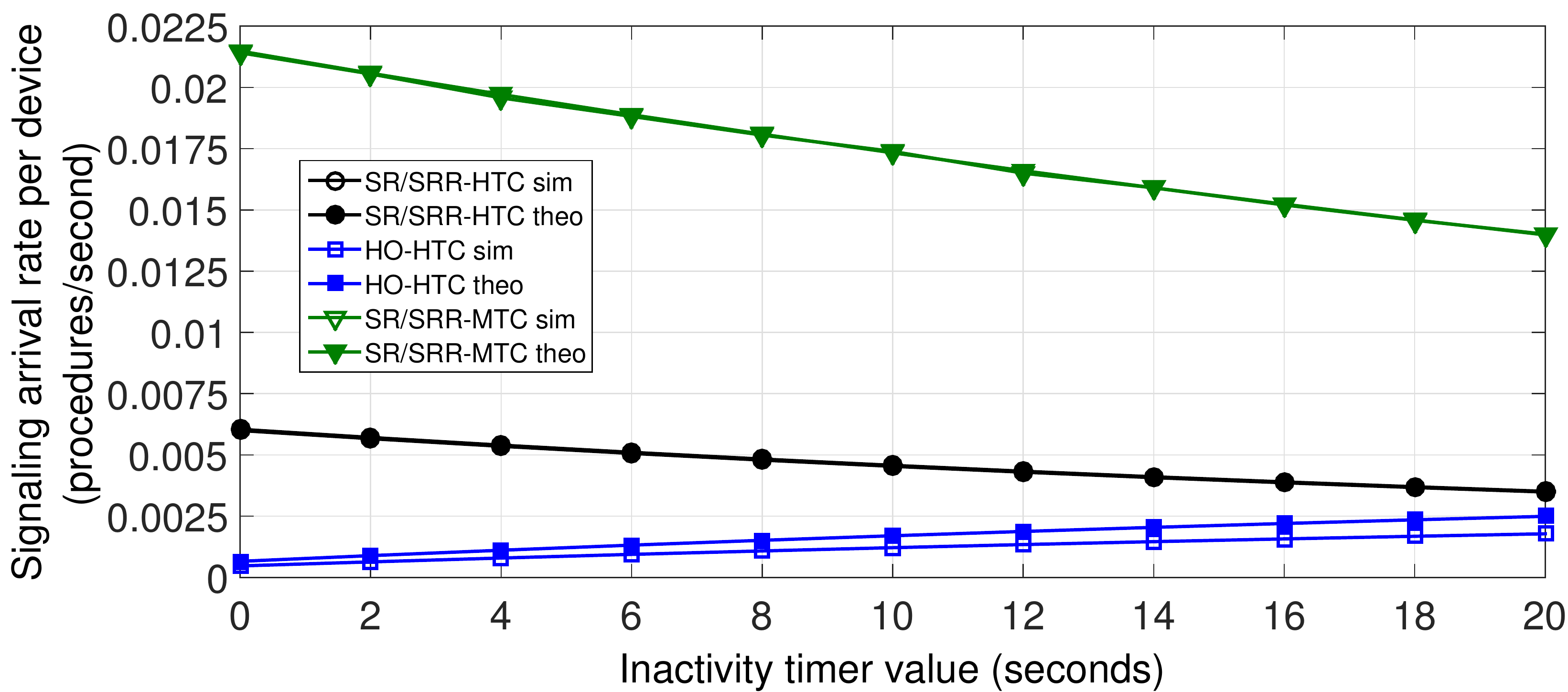}
\end{center}
\caption{Control procedures arrival rates versus user inactivity timer.}
\label{fig:lambdas}
\end{figure}

To characterize the control messages arrival rate at the vMME, we generated signaling traces for 20000 UEs and 20000 MTCDs for several \emph{inactivity timer} $(T_I)$ values.

The obtained results are depicted in  Fig. \ref{fig:lambdas}. It shows the mean arrival rates for the different signaling procedures, by using the provided theoretical model -referred to as label \emph{theo}- and after the conducted simulations -referred to as label \emph{sim}- as a function of the inactivity timer $T_I$.  

Results show that the SRs and SRRs rates decrease with $T_I$ for both HTC and MTC traffics. One possible explanation is because the higher the value of the inactivity timer, the smaller the probability the timer runs out within an inter AAP. Thus, the user stays 
 in the connected state between consecutive AAPs, avoiding the need for triggering procedures to reserve and release resources.
Conversely, the HRs rate slightly increases with the timer value, since in these cases the user remains connected longer after an AAP. Consequently, there is a higher chance that a user will be in connected state when a cell crossing event takes place.

Please note that the amount of signaling traffic generated by MTCs is higher than for HTCs case. From Fig. \ref{fig:lambdas}, it is observed that $\lambda_{U}^{SR}=0.0045$, $\lambda_{U}^{HR}=0.0012$ and $\lambda_{S}^{SR}=0.0173$ 
 procedures per second and terminal for $T_{I}=10s$. That means each MTC device generates about $3.5$ times more control messages than an HTC UE.

For the schemes depicted in Fig. \ref{fig:lambdas},  Table \ref{tab:mse_lambdas} shows the root-mean-square error (RMSE) between the experimental rates (obtained after simulations) and predicted ones (using the proposed model). It demonstrates that the analytic expressions
(e.g., Equations \ref{eq:lambdaSR}, \ref{eq:lambdaHR}, and \ref{eq:lambda_sr_mtc}) fit the experimental data obtained by simulation.


\begin{table}[t]
\renewcommand{\arraystretch}{1.5}
\centering
\caption{RMSE for predicted arrival rate per device (see Fig. \ref{fig:lambdas}).}
\label{tab:mse_lambdas}
\begin{tabular}{|c|c|c|}
\hline
$RMSE(\lambda_{U}^{SR})$ & $RMSE(\lambda_{U}^{HR})$ & $RMSE(\lambda_{S}^{SR})$ \\ \hline
$4.07\cdot 10^{-5}$      & $15.0\cdot 10^{-4}$      & $6.65\cdot 10^{-5}$      \\ \hline
\end{tabular}
\end{table}

The higher prediction error for the HR procedure rate is due to the fluid-flow mobility model implementation: a bounce-back strategy is employed when a user reaches an edge of the geographical area. That decreases the $\overline{r}_{cc}$ per user in comparison with the predicted by the fluid flow model expression.       

\subsection{System Delay}

\begin{figure*}
\center{
\subfigure[One MTC device per each UE (Scenario 1).]{\label{fig:a}\includegraphics[width=\columnwidth]{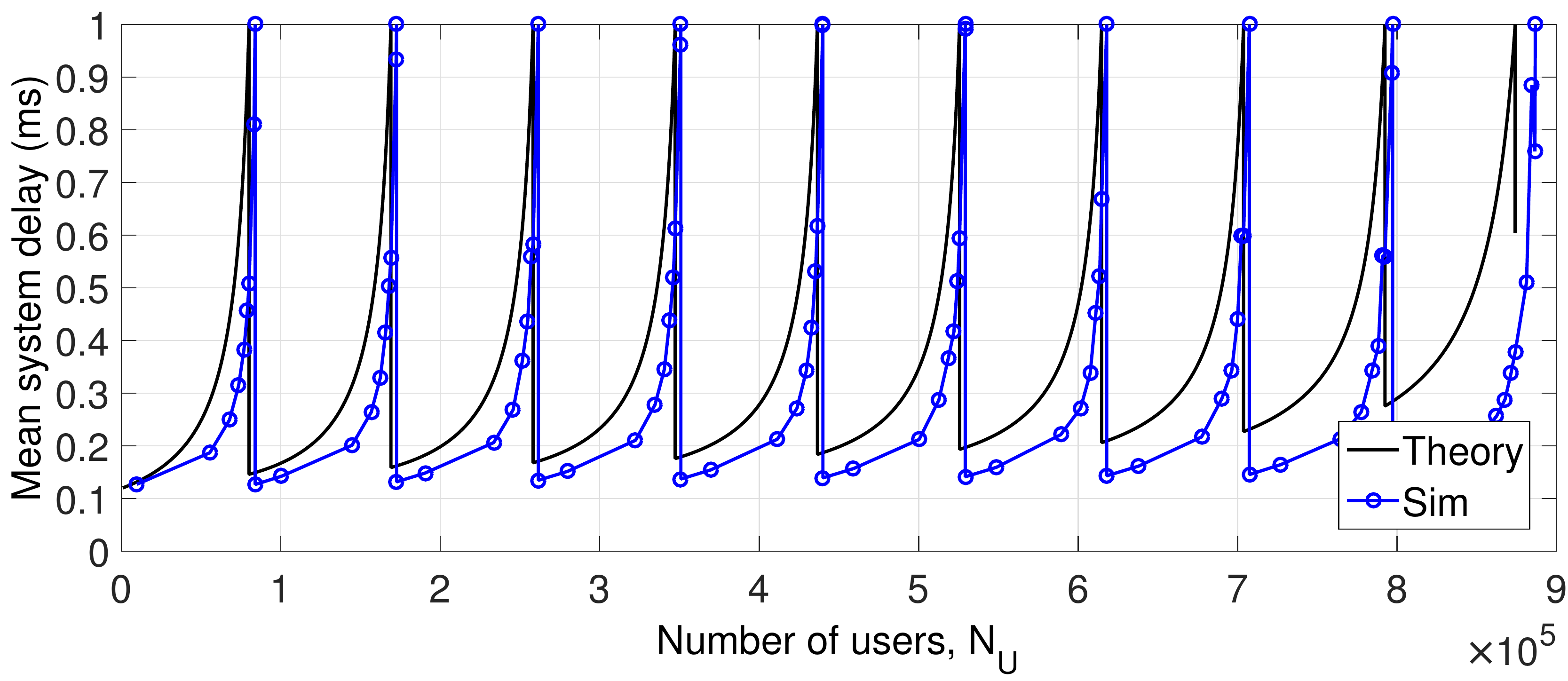}}
\subfigure[Three MTC devices per each UE (Scenario 2).]{\label{fig:b}\includegraphics[width=\columnwidth]{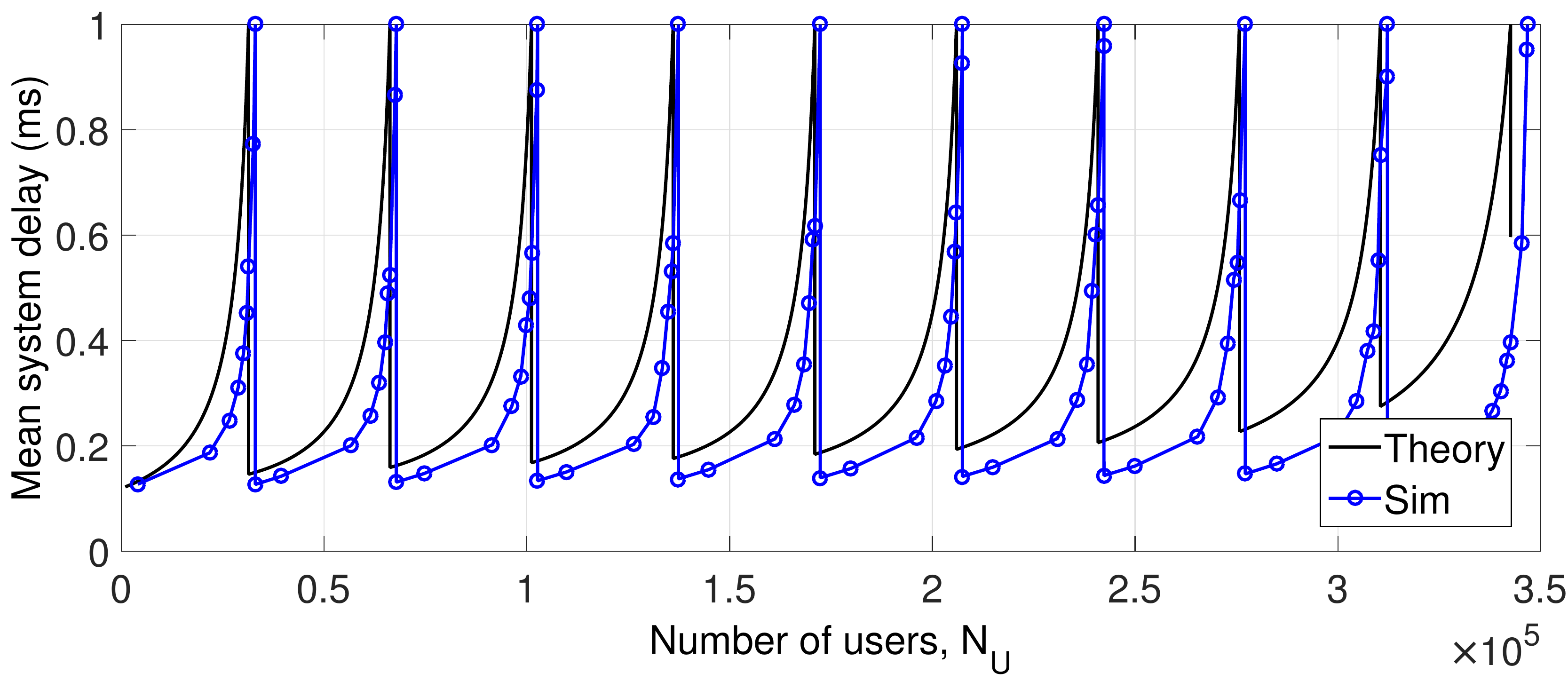}}
}
\caption{Mean system delay vs number of users.}
\label{fig:system-delay}
\end{figure*}

Most mobile networks standards requirements define a delay budget to perform the control procedures. 
 In order to evaluate the delay of our system and to compare the experimental results with the theoretical ones, we generated a signaling trace for $1200000$ UEs and a $T_I=10s$. 
 
 Two scenarios were considered: i) one with one MTC device per each UE (referred to as \emph{scenario 1}), and ii) the other with three MTC devices per each UE (referred to as \emph{scenario 2}). 
 
 Figs. \ref{fig:system-delay} depict the mean system delay versus the number of users for both theory (Equation \ref{eq:T}) and simulation for the two scenarios. The curves were generated using the dimensioning algorithm introduced in subsection \ref{subsec:dimensioning-criterion-theo} for $T_{max}=1\,ms$. 
 
 As a general trend, given a number of MME SLs, the system delay grows with the number of users. There is a point where the number of MME SL instances cannot withstand the control messages arrival rate and the system delay shoots up. When $T=T_{max}$, in order to limit the system delay, a new MME SL must be instantiated to cope with the control plane workload. This fact explains the periodic spiky pattern of Fig. \ref{fig:system-delay}. 
 

Visibly, though simulation and theoretical results show a similar shape, delays are smaller in case of the simulation. This is due to assumptions of the theoretical system differ from those adopted in our simulation model implementation. For instance, the simulation model considers deterministic service rates. 

The root-mean-square error between simulation and theoretical results are $0.34$ and $0.35$ ms for scenario 1 and 2, respectively. Notably, it can be observed that the error increases with $m$ (the number of MME SLs). That trend can be explained because  the theoretical delay impact contribution of the database 
begins to be noticeable, earlier than in the simulation setup.
\subsection{vMME Dimensioning}
\label{sec:nrControllerDimensioning}


\begin{figure*}
\center{
\subfigure[{vMME capacity versus the number of MME SL instances for both scenarios 1 and 2.}]{\label{fig:vMMEdim_a}\includegraphics[width=\columnwidth]{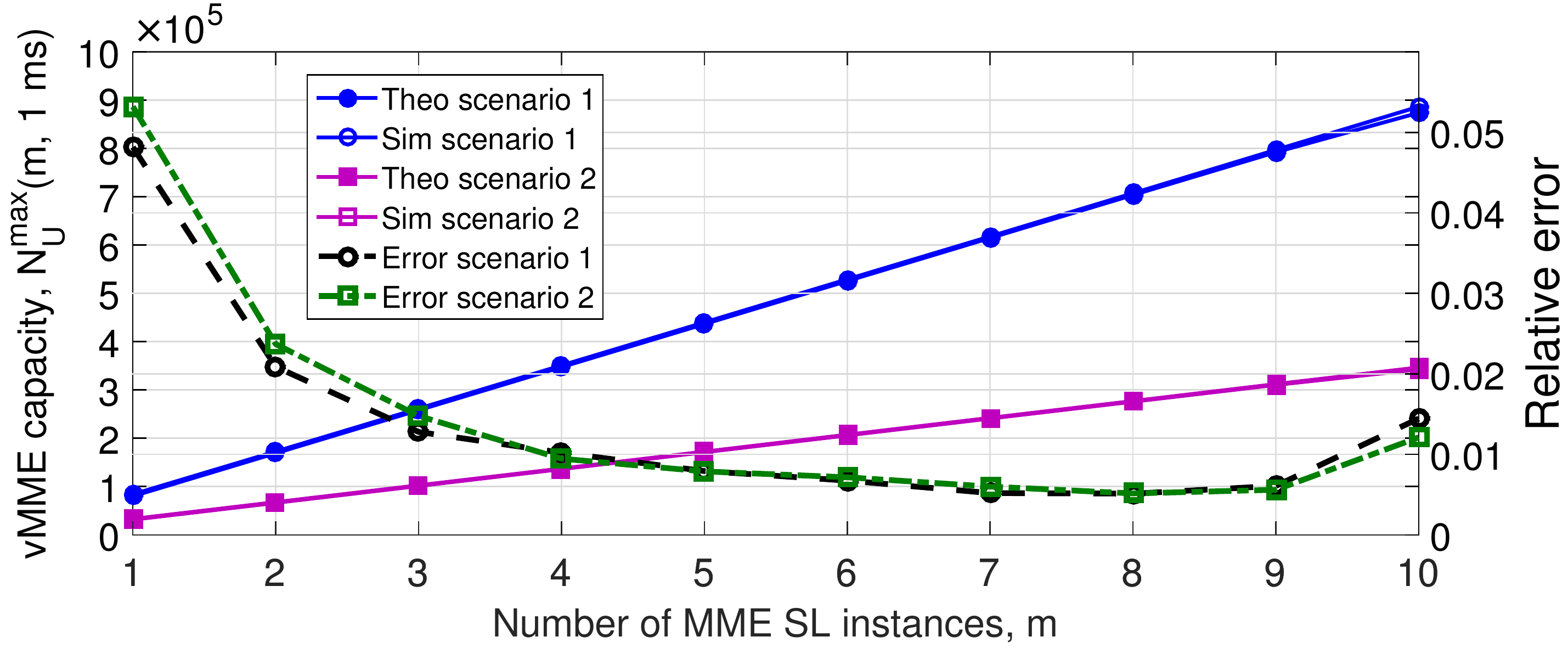}}
\subfigure[{vMME capacity versus the number of MME SL instances for different UE speeds (Scenario 1).}]{\label{fig:vMMEdim_b}\includegraphics[width=\columnwidth]{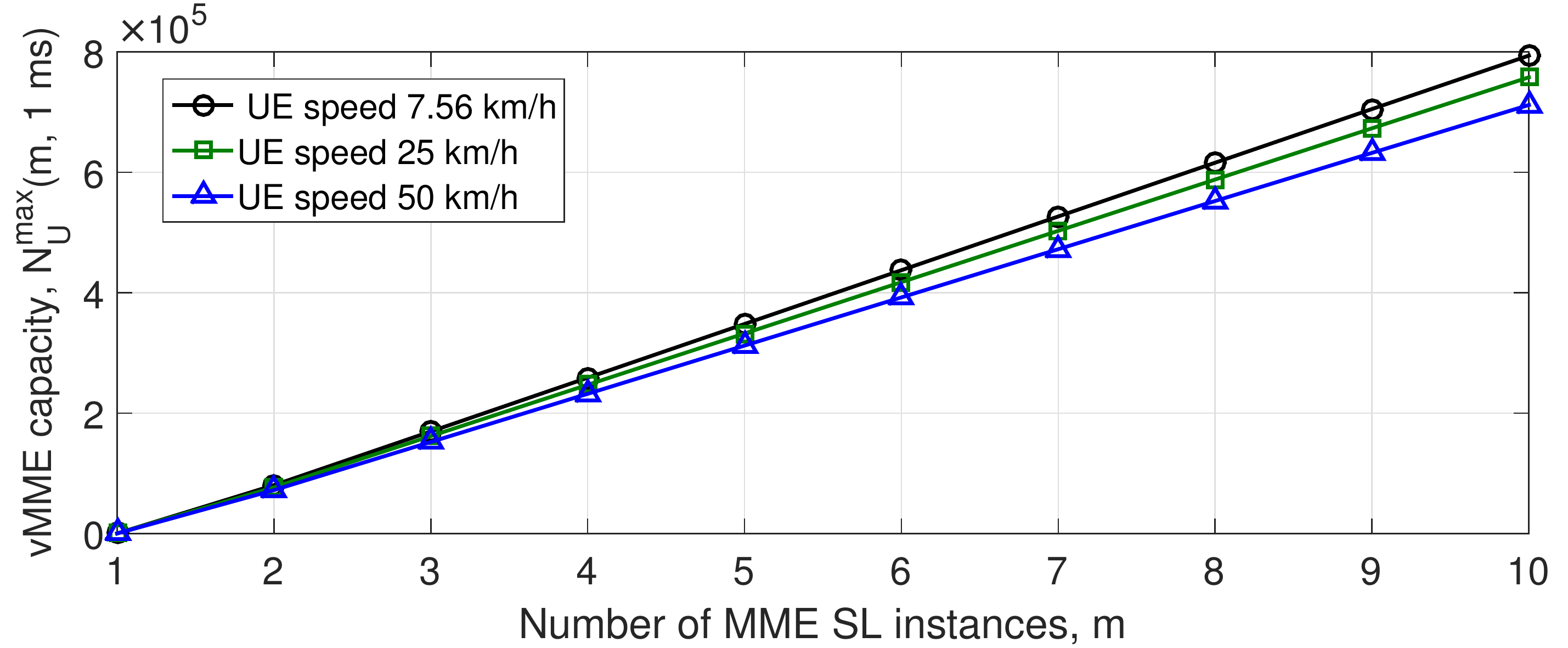}}
}
\caption{vMME capacity.}
\label{fig:system-delay}
\end{figure*}


A major application of the proposed theoretical framework is vMME dimensioning. That is, to predict the minimum number of MME SL instances required to guarantee $\overline{T}=\overline{T}_{max}$ given the number of UEs and MTC devices. 

Let $N_{U}^{max}(m, \overline{T}_{max})$ denote the maximum number of UEs supported by the vMME (i.e., vMME capacity) depending on the number of SL instances $m$ and the target mean system delay $\overline{T}_{max}$. To assess the 
goodness of our mathematical model, we computed $N_{U}^{max}(m, \overline{T}_{max})$ versus the number of MME SL instances $m$ for $T_{max}=1$ ms for both scenarios 1 and 2 (see Fig.  \ref{fig:vMMEdim_a}).
The relative error between the theoretical and experimental curves ranges roughly from $0.5\%$ to $5.5\%$. Therefore, we can conclude that the mathematical model is useful for dimensioning purposes. Furthermore, in general, the error decreases as $m$ increases, except for $m=10$. As it was mentioned in the previous section, this fact can be explained because the theoretical delay impact contribution of the database starts to be remarkable, earlier than in the simulation setup. 

	Since $\overline{T}_{max}$ might depend on the Quality of Service (QoS) requirements of the service and scenario considered, we assessed the vMME capacity $N_{U}^{max}(m, \overline{T}_{max})$ for different values of $m$ and $\overline{T}_{max}$. Specifically, the considered ranges for $\overline{T}_{max}$ and $m$ were 500 $\mu s$ to 3 ms and 1 to 10 instances, respectively. The results obtained show that the vMME capacity does not differ significantly in the range of values considered for $\overline{T}_{max}$. For $\overline{T}_{max}=500\,\mu s$  vMME capacity decreases 0.94\% and 1.25\% in comparison with $\overline{T}_{max}=3\,ms$ case for Scenarios 1 and 2, respectively. 

Finally, we also assessed the impact of average user mobility speed $\overline{v}$ on the vMME capacity (see Fig. \ref{fig:vMMEdim_b}). As we assumed MTCs without mobility, we have considered Scenario 1 because in it the number of sensors per UE is lower. Thus, we can appreciate better the effects of mobility on MME capacity. Given that we assumed a fluid-flow model, the relationship between $\lambda_{U}^{HR}$ and $\overline{v}$ is linear. Therefore, for the speeds considered $\overline{v} = 7.56\,km/h$, $\overline{v} = 25\,km/h$, and $\overline{v} = 50\,km/h$ the average handover rates per user are $\lambda_{U}^{HR}=0.0012$, $\lambda_{U}^{HR}=0.0040$, and $\lambda_{U}^{HR}=0.0080$ procedures per second, respectively. We obtained that the capacity of the vMME decreases 6.26\% for a doubling of $\overline{v}$.


%
\subsection{Scalability Analysis}
\label{sec:salabilityanalysis}

 The scalability assessment of the system modeled depends on the exact cost function of the supporting cloud service.
 
 As an example, we consider the Amazon EC2 Service, with the costs and configuration detailed in Table \ref{tab:ec2-configuration}. We assume a medium sized CPU instance \emph{m3.xlarge} with an average $11.38\cdot 10^9$ float operations per second \cite{iosup2011performance}. Our setup also includes the  Amazon Aurora database \cite{aurora-benchmark-2015}, which is reported to provide $10^5$ updates/s transactions. The target mean system delay is set $\overline{T}_{max}=1\,ms$. 
  
\begin{table*}[tb]
\centering
\caption{Cloud service configuration and cost calculation.}
{\footnotesize  
\begin{tabular}{l l l}
\hline
    Cost & Configuration & Calculation\\ 
\hline
\hline
$C_{ci_{type}}(k)$ & \emph{m3.xlarge} instance rental ($0.266\$$/hour) &  $0.266/3600$ \\ 
$C_{ci_{stor}}(k)$ &   Local storage per month (10GB), and optimized data access ($0.025\$$/hour). &  $10 \cdot 0.10 + 0.025 / 3600$ \\ 
$C_{ci_{thro}}(k)$ & Supposing $I_{size}=200$ bytes, this is the data sent from the datacenter, calculated as $\lambda \cdot 200$ & \begin{tabular}{@{}r l@{}}0.000(\$)/GB & First GB/month \\ 0.090(\$)/GB & Up to 10 TB/month\\  0.085(\$)/GB & Next 40 TB/month\\ 0.070(\$)/GB & Next 100 TB/month\\ 0.050(\$)/GB & Next 350 TB/month\\ \end{tabular} \\
$C_{db_{type}}(k)$& Aurora $db.r3.8xlarge$ instance ($4.64\$$/hour) & $4.64/3600$\\
$C_{db_{stor}}(k)$& $0.1\$$ per GB/month, for a total database size of $N_U \cdot 1KB$. & $(0.1  \cdot N_U \cdot 1024\cdot \lambda/1e9 )/ 2628000$ \\
$C_{db_{thro}}(k)$& $0.2\$$ per million transactions/month & $ 0.2 \cdot \lambda /1e6  $\\
$C_{b_{type}}(k)$& Service fee of $0.025\$$/month & $0.025 / 2628000$\\
$C_{b_{thro}}(k)$& $0.008\$$ per GB serviced, supposing $O_{size}=200$ Bytes & $\lambda \cdot 0.008 \cdot 200/1e9$\\
\hline
\end{tabular}
}
\label{tab:ec2-configuration}
\end{table*} 

\begin{figure}[tb]
\begin{center}
\includegraphics[width=1.0\columnwidth]{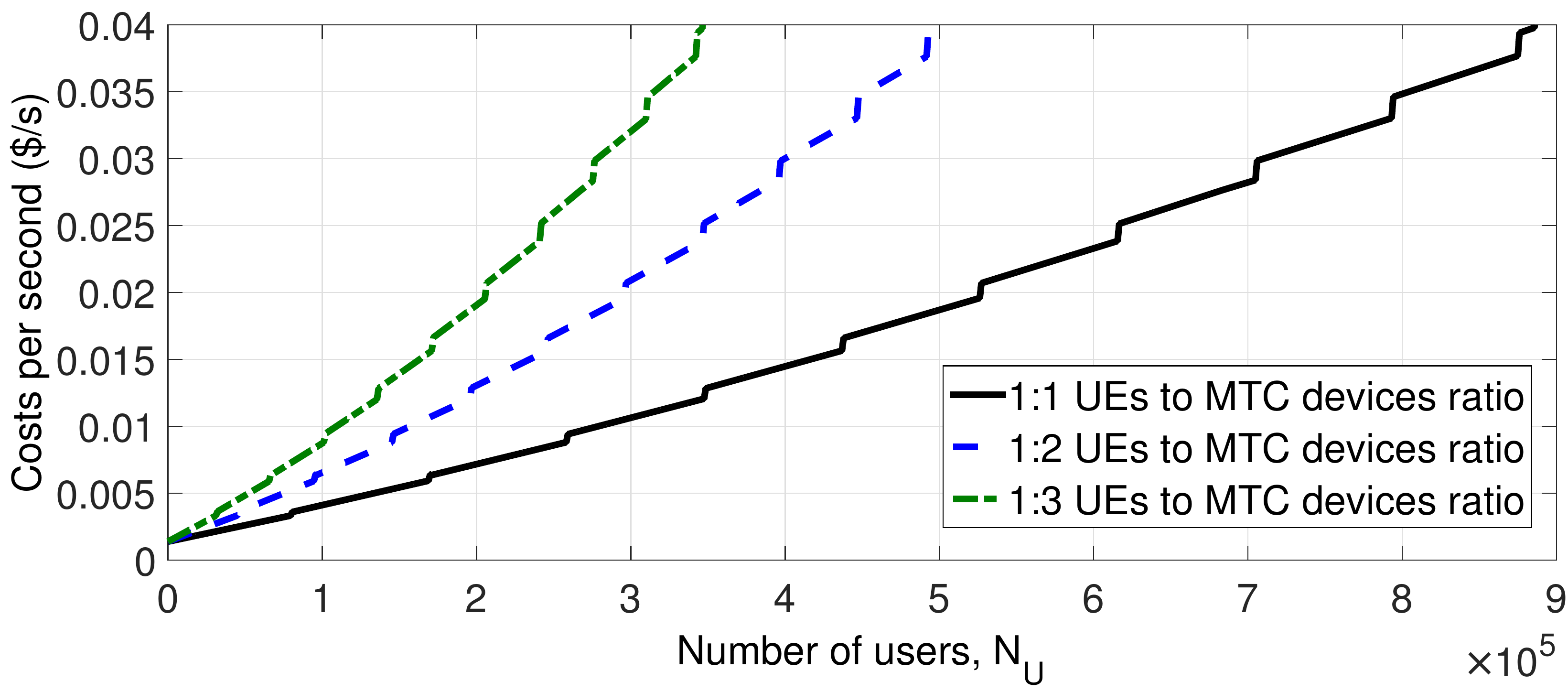}
\end{center}
\caption{Cost per second vs number of users.}
\label{fig:costs}
\end{figure}

Assuming an on demand cloud service, Fig. \ref{fig:costs} depicts the running costs of the system (measured in \$/s) for the selected configuration. It includes three scenarios for different UE to MTC device ratio.
Interestingly, in general, it shows that the running cost of the virtualized vMMEs is almost linear with the number of users in the system. Nevertheless, note that the overhead costs of deploying
new instances hinder the system scalability regarding the number of MME SL instances.

\begin{figure}[tb]
\begin{center}
\includegraphics[width=1.0\columnwidth]{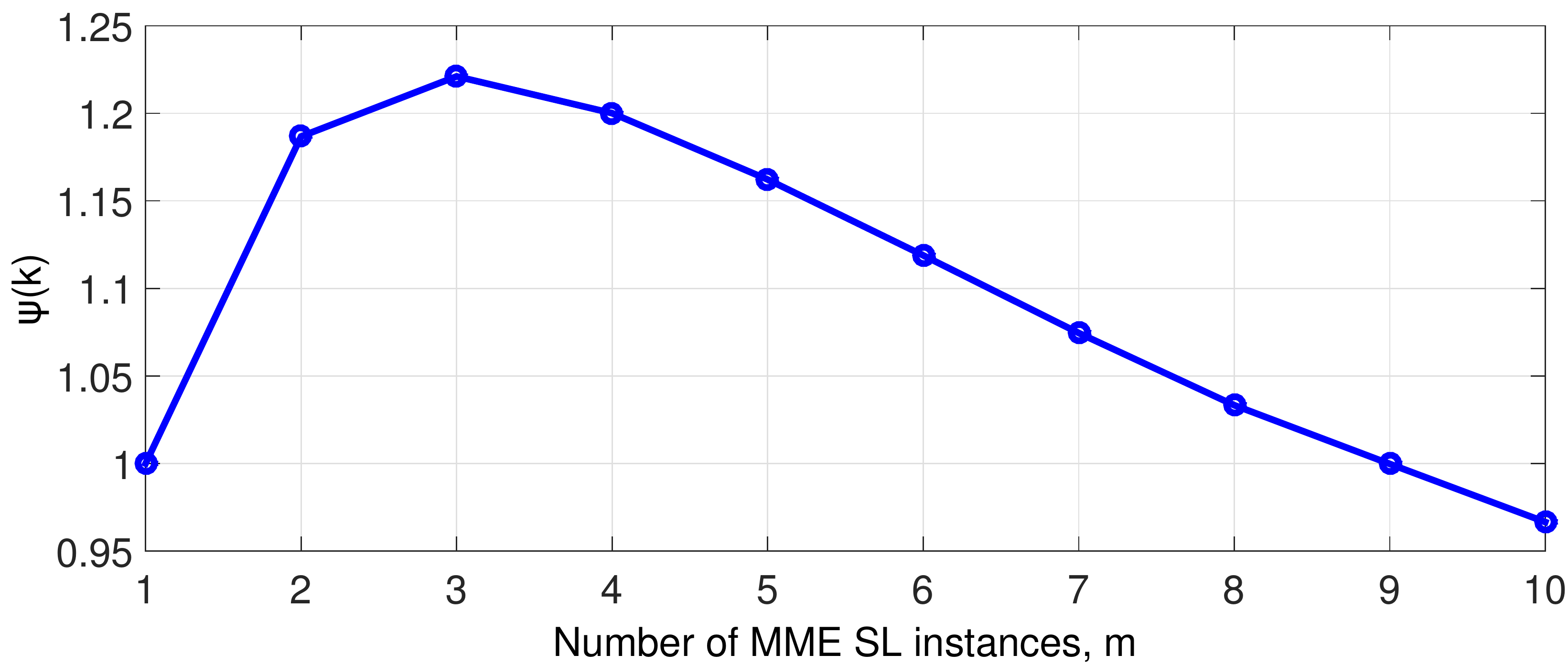}
\end{center}
\caption{Scalability $\Psi(k)$ vs number of MME SL instances.}
\label{fig:scalability}
\end{figure}

Finally, for testing the scalability of the proposed virtualized system in Fig. \ref{fig:scalability} the scalability metric $\psi(k)$ (Section \ref{sec:Scalability-metrics}) is depicted as a function of the number of MME SL instances.

Interestingly, for the configuration considered, the system is positively scalable regarding the number of MME SL instances for $m<10$. 
However, beyond that point, it holds that $\psi(k)<1$; in other words for more than 9 MME SL instances, the system is not perfectly scalable. This limit stem from the database utilization reaches about 100\% of its capacity. At that point, it would be necessary to scale up the database. 

Nevertheless, recall from Fig. \ref{fig:vMMEdim_a} that the system can serve roughly $900000$ UEs and $900000$ sensors for Scenario 1 and more than $325000$ UEs and $3\cdot 325000$ MTCDs for Scenario 2 (i.e., this is the equivalent of a signaling workload around 37000 LTE control procedures per second) for $\overline{T}_{max}=1\,ms$ and $m=10$, where the MME still scales positively. The vMME capacity obtained is in the same order of magnitude as non-virtualized MME solutions \cite{sgsn-mme-ericsson}.

\section{Conclusions and Future Work} 
\label{sec:conclusions}

  In this paper, we propose a theoretical framework  to assess the system response time of a vMME running in a data center. This framework includes: i) a queuing network to model a vMME in a data center, and ii) expressions to estimate the rates of different signaling procedures (e.g., SR, SRR, and HR) per UE and MTC device, as well as to estimate the overall system delay. We have validated this framework by simulation. Additionally, we have estimated the processing time of a vMME for the different types of LTE control procedures considered. 
  
Using this framework, a vMME dimensioning procedure is provided to predict the number of MME SL instances given a system delay budget ($T_{max}$), the number of UEs, and the number of MTC devices. After the conducted experiments, results show that our mathematical model is accurate for that purpose. Specifically, we have obtained a relative error between the theoretical and simulation results below $5.5\%$. 
  
  
  Finally, we have carried out a scalability analysis of the system. The reported results for the typical configuration considered suggest that MME virtualization is scalable for signaling workloads up to 37000 control procedures per second considering a data center with commodity hardware. This limit stem from the database utilization reaches its maximum. In order to continue the scalability analysis beyond this point, it would have to consider a strategy to scale the state database.

Regarding the future work, several challenges lie ahead. One of the main challenges is the design of a dynamic capacity provisioning algorithm for the vMME. This is to scale up or down the resources (e.g., vMME SL instances) assigned to the vMME depending on the fluctuations in signaling workload. Mobile network traffic exhibits long-term variations such as time-of-day or seasonal effects, as well as short-term fluctuations caused by unexpected events. 

\section*{Acknowledgment}
This work is partially supported by the Spanish Ministry of Economy and Competitiveness and the European Regional Development Fund (project TIN2013-46223-P), and the Spanish Ministry of Education, Culture and Sport (FPU grant 13/04833).

\bibliographystyle{IEEEtran}
\bibliography{references} 
%
%
%

%

\ifieeetran
\ifbiography
	\begin{IEEEbiography}[{\includegraphics[width=1in,height=1.25in,clip,keepaspectratio]{jorgenavarro}}]{Jorge Navarro-Ortiz}
	received the M.Sc. in Telecommunications Engineering from the University of Málaga (Spain) in 2001. After working at Nokia Networks, Optimi (Ericsson) and Siemens, he joined the Department of Signal Theory, Telematics and Communications of the University of Granada as Assistant Professor in 2006, where he finished his Ph.D. theses in 2010 with the European Doctorate mention after performing a research stay at the Dipartimento di Ingegneria dell'Informazione of the University of Pisa. Since June until September 2012 he was a visiting researcher with the BWN lab of the School of Electrical and Computer Engineering of the Georgia Institute of Technology. His research interests include cognitive radio, heterogeneous networks, LTE and LTE-Advanced systems among others.
	\end{IEEEbiography}

	\begin{IEEEbiography}[{\includegraphics[width=1in,height=1.25in,clip,keepaspectratio]{pabloameigeiras}}]{Pablo Ameigeiras}
	received his M.Sc.E.E. degree in 1999 by the University of Málaga, Spain. He carried his Master Thesis at the Chair of Communication Networks, Aachen University (Germany). In 2000, he joint the Cellular System group at the Aalborg University (Denmark) where he carried out his Ph.D. thesis. After finishing his Ph.D., Pablo worked in Optimi/Ericsson. In 2006, he joined the Department of Signal Theory, Telematics, and Communications at the University of Granada (Spain). Since June until September 2012 he was a visiting researcher with the BWN lab of the School of Electrical and Computer Engineering of the Georgia Institute of Technology. Currently, his research interests include 3G LTE and LTE-Advanced systems.
	\end{IEEEbiography}

\fi
\fi







\end{document}